\documentclass[fleqn,10pt]{wlscirep}

\usepackage[utf8]{inputenc}
\usepackage[T1]{fontenc}
\usepackage[english]{babel}
\usepackage{float}
\usepackage{graphicx}
\usepackage{upgreek}
\usepackage{amssymb}
\usepackage[normalem]{ulem}
\usepackage{booktabs}
\usepackage{multirow}
\usepackage{subcaption}

\newcommand\blfootnote[1]{%
  \begingroup
  \renewcommand\thefootnote{}\footnote{#1}%
  \addtocounter{footnote}{-1}%
  \endgroup
}

\title{Local insulator-to-superconductor transition in amorphous InO$_x$ films modulated by e-beam irradiation}

\author[1,2,3,*]{Iago F. Llovo}
\author[4,$\dagger$]{Julien Delahaye}
\affil[1]{Centro de Supercomputación de Galicia (CESGA), 15705 Santiago de Compostela, Spain}
\affil[2]{QMatterPhotonics, Departamento de Física de Particulas, Universidade de Santiago de Compostela, 15782 Santiago de Compostela, Spain}
\affil[3]{iMATUS, Universidade de Santiago de Compostela, 15782 Santiago de Compostela, Spain}
\affil[4]{Univ. Grenoble Alpes, CNRS, Grenoble INP, Institut Néel, 38000 Grenoble, France}

\begin{abstract}
We present a novel method enabling precise post-fabrication modulation of the electrical resistance in micrometer-scale regions of amorphous indium oxide (a-InO$_x$) films. By subjecting initially insulating films to an electron beam at room temperature, we demonstrate that the exposed region of the films becomes superconducting. The resultant superconducting transition temperature ($T_c$) is adjustable up to 2.8~K by changing the electron dose and accelerating voltage. This technique offers a compelling alternative to traditional a-InO$_x$ annealing methods for both fundamental investigations and practical applications. Moreover, it empowers independent adjustment of electrical properties across initially identical a-InO$_x$ samples on the same substrate, facilitating the creation of superconducting microstructures with precise $T_c$ control at the micrometer scale. Some possible mechanisms for the observed resistance modifications are discussed.
\end{abstract}

\begin{document}

\flushbottom
\maketitle
\thispagestyle{empty}

\begin{figure}[t]
    \centering
    \includegraphics[width=0.75\textwidth]{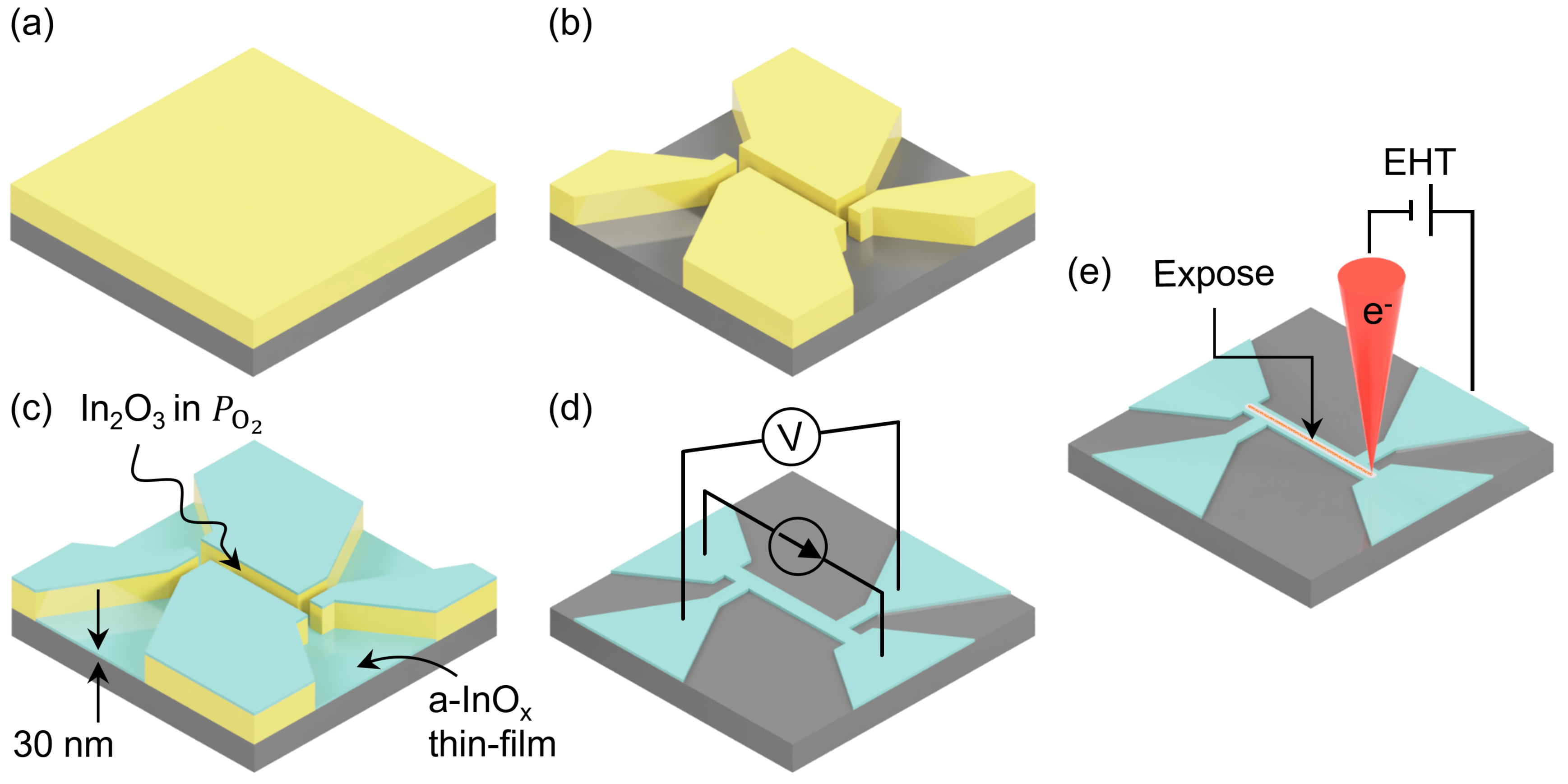}
    \vspace{10pt}
    \caption{(a) A negative PMMA resist mask (yellow) was deposited on highly-doped Si substrates coated with 100~nm of thermally grown SiO$_2$ and (b) patterned using standard EBL. (c) a-InOx thin films (cyan) were then deposited in an oxygen-controlled atmosphere to obtain samples close to the SIT. (d) After lift-off, the pristine samples evolved at room temperature for over a day and were characterized by measuring the $R(T)$ curves down to $\sim1.6$~K. (e) Controlled e-beam exposure to pattern microstructures on the InOx channels resulted in increased conductivity, eventually crossing the SIT.}
    \label{Figure1}
\end{figure}

\begin{figure}[!htbp]
    \centering
    \includegraphics[width=0.62\linewidth]{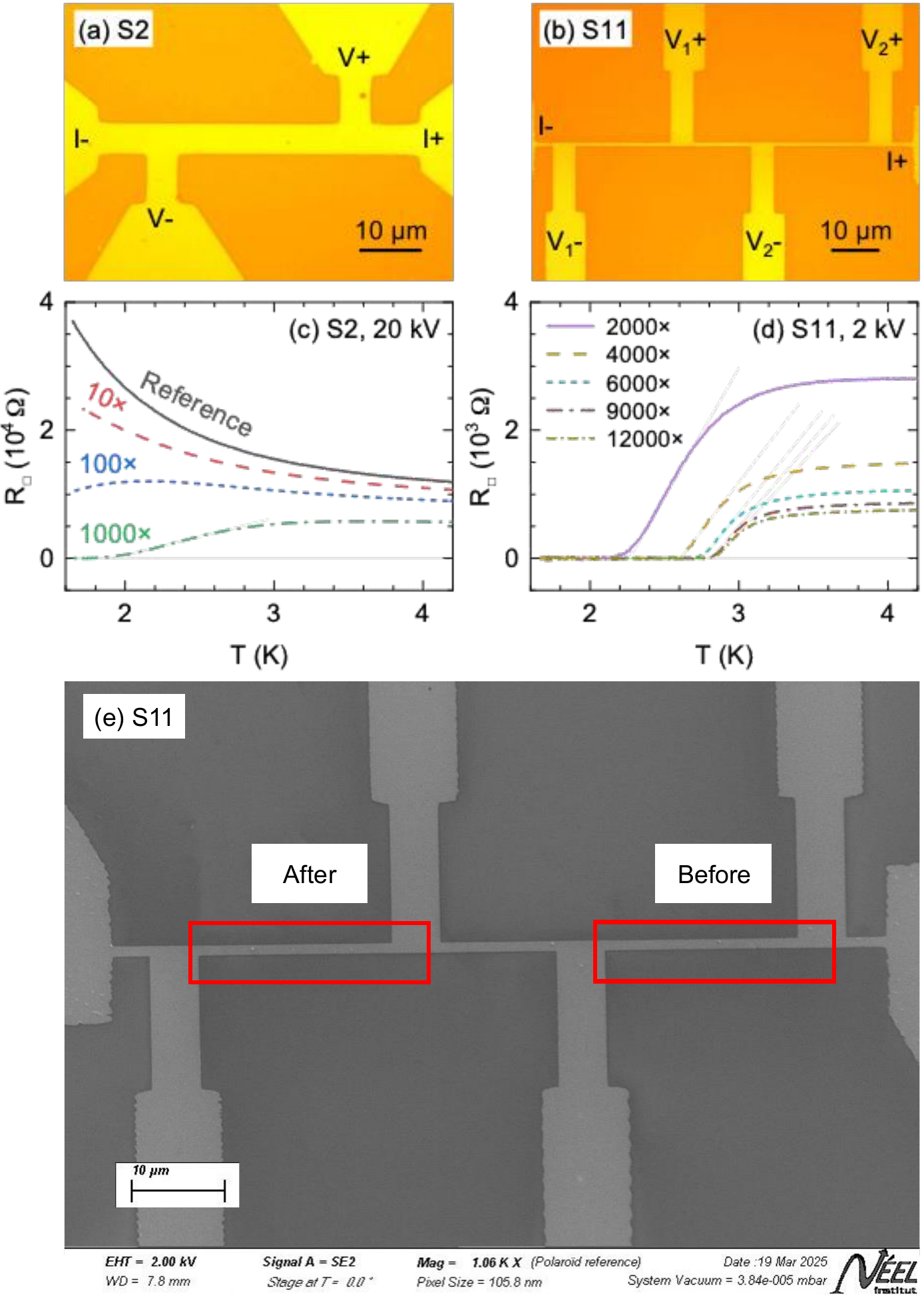}
    \caption{(a,~b) Geometry of two channels from substrates S2 and S11, and (c,~d) their corresponding $R_\square (T)$ curves before (reference) and after the e-beam treatment. Doses are indicated as multiples of the standard dose for PMMA resist, 250~$\upmu$C~cm$^{-2}$. A 20~kV e-beam was used to expose samples on substrate S2, finding a progressive evolution with dose, from insulating to superconducting behavior. With a dose of 1000$\times$, the sample became superconducting below 1.9~K. A 2~kV e-beam at higher doses was used to expose samples on S11. All the channels of sample S11 became superconducting after the treatment, with a progressive increase of $T_c$ with dose, from 2.28~K up to 2.84~K, and a diminution of the normal state sheet resistance from 12~k$\Omega$ at 4.2~K (reference sample, not shown) down to less than 1~k$\Omega$ for a dose of 12000$\times$. This anticorrelation of normal state resistivity and $T_c$ is also present in a-InO$_x$ samples made by tuning the O$_2$ pressure during deposition. (e) SEM micrograph of S11, where the left channel had been previously treated (12000$\times$ dose) and the right one was kept pristine until before the capture. No morphological changes could be observed between untreated and treated channels by visual inspection.}
    \label{Figure2}
\end{figure}

\section*{Introduction}

\blfootnote{
\normalsize $^*$ ifllovo@cesga.es (Corresponding author)}
\blfootnote{
\normalsize $^\dagger$ julien.delahaye@neel.cnrs.fr}
Disordered amorphous indium oxide (a-InO$_x$) films have long served as an experimental platform for studying the electrical properties of disordered systems. During the 1980s, many theoretical predictions related to the Mott-Anderson metal-insulator transition were verified in this system \cite{OvadyahuSS1982, ImryPRL1982, OvadyahuJPC1986} and in the early 1990s, two new topics rose from disordered indium oxide films measurements. The first topic was the search for an electron glass \cite{DaviesPRL1982, GrunewaldJPCS1982, PollakSEM1982}, with the reports of slow and glassy relaxations of the conductance in field effect experiments on insulating films \cite{BenChorinPRB1993, VakninPRB2002, PollakBookElectronGlass2012}. The second topic was the study of the superconducting-insulator transition (SIT) in amorphous films as a function of magnetic field \cite{HebardPRL1990, GantmakherJETPL2000, SambandamurthyPRL2004} and disorder \cite{ShaharPRB1992}. The nature of this transition, especially the insulating state terminating superconductivity, remain the subject of heated experimental and theoretical debates \cite{SambandamurthyPRL2005, SacepeNP2020}. The unusual properties of this insulating state may indeed reflect the existence of a many body localized phase \cite{OvadiaSR2015}, a topical and striking issue in condensed matter physics \cite{GornyPRL2005, BaskoAP2006, OganesyanPRB2007, NandkishoreARCM2015, AletCRP2018}.

From an applications perspective, a-InO$_x$ films are also a promising system for the fabrication of high kinetic inductances. Such electrical circuit elements can be obtained from superconducting a-InO$_x$ films approaching the SIT, in the form of long wires of a few micrometers wide \cite{AstafievNature2012, Charpentier2024} from which low-loss and small size microwave resonators can be made, crucial for the development of quantum circuits \cite{AstafievNature2012, SacepeNP2020, Charpentier2024}. Additionally, these films have been used to manufacture sensitive kinetic inductance photon detectors \cite{DupreSST2017}. Doped a-InO$_x$ films have also found extensive use in the production of semiconducting microdevices, such as transparent thin-film transistors (TFT). As oxygen-vacancies in a-InO$_x$ exhibit electronic properties equivalent to those of tin-doped amorphous indium-tin oxide (a-ITO) \cite{Bellignham91}, fields already benefiting from the commercial use of a-ITO (e.g.,transparent display research) could also potentially leverage the control of the electronic properties of a-InO$_x$. 

The prospects for fundamental studies and applications of a-InO$_x$ rely on the ability to control and adjust the electrical properties of these films, which exhibit significant variability across the SIT. These films are typically prepared by sputtering or electron beam (e-beam) deposition of pure In$_2$O$_3$ under a partial pressure of oxygen \cite{OvadyahuJPC1986, PashmakovJNCS1993, PashmakovSSC1993, FritzscheSEM1994, ClaflinJEM1996, GivanPRB2012}. Oxygen vacancies are created during deposition, so the stoichiometric composition of the amorphous film is less oxygen-rich than the In$_2$O$_3$ target, and can be regulated by the relative oxygen pressure during deposition \cite{Dawar1984}. Additionally, the deposition conditions can also affect the crystallinity of the thin films: microcrystalline films result from evaporation at substrate temperatures exceeding 150$^\circ$C, while substrate temperatures below 40$^\circ$C yield amorphous films \cite{OvadyahuJPC1986}. Therefore, common methods for adjusting the electrical properties of the resulting films include adjusting the oxygen pressure, the evaporation rate or the film thickness during the evaporation process, and employing post-evaporation annealing \cite{ShaharPRB1992, SacepePRB2015}. In the case of amorphous films, annealing steps restricted to 40--60$^\circ$C lead to irreversible reductions in electrical resistance without inducing crystallization \cite{OvadyahuJPC1986, ShaharPRB1992}.

In this article, we introduce a novel technique for modulating the electrical properties of a-InO$_x$ films in relation to the SIT. We demonstrate that upon exposure to the e-beam of a scanning electron microscope (SEM), an initially insulating film undergoes a local transformation into the superconducting phase across the SIT, featuring a superconducting transition temperature of up to 2.8~K. Moreover, we demonstrate the capability of controlled e-beam exposure to locally manipulate the SIT, achieving a lateral resolution of a few micrometers. Finally, we discuss the potential origins of these alterations and the opportunities afforded by this new technique.

\section*{Methods}

\begin{figure}[t]
    \centering
    \includegraphics[width=0.6\textwidth]{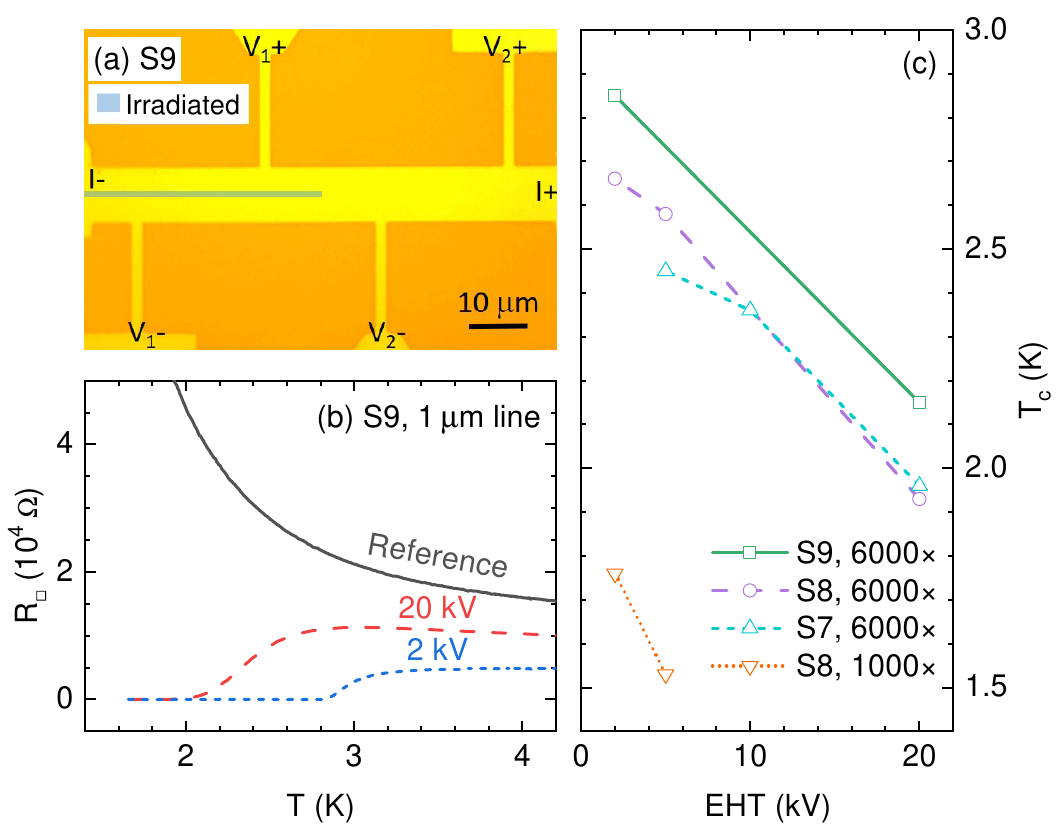}
    \caption{Results of the e-beam treatment with different acceleration voltages. (a) Diagram of the irradiated area, a $1~\upmu$m wide line in the middle of the a-InO$_x$ channel, represented as the shaded area on top of the optical microscope image. (b) $R_\square (T)$ curves measured for substrate S9 before (reference, solid black line) and after the treatment. A clear $T_c$ increase was observed when the acceleration voltage of the electrons was reduced from 20~kV (long-dashed red line, $T_c = 2.0$~K) to 2~kV (short-dashed blue line, $T_c = 2.8$~K). (c) Evolution of $T_c$ with the acceleration voltage EHT for three different samples.}
    \label{Figure3}
\end{figure}

\subsection*{Sample deposition}
The sample fabrication and treatment procedures are illustrated in Fig.~\ref{Figure1}. Firstly, the insulating a-InO$_x$ films were manufactured via e-beam evaporation. A resist mask was first made using standard e-beam lithography (EBL) on highly-doped Si substrates coated with 100~nm of thermally grown SiO$_2$. Subsequently, pure In$_2$O$_3$ was e-beam evaporated and deposited on the masked substrates at a rate of 2~\AA/s under a partial O$_2$ pressure. The evaporation was stopped after depositing 30~nm on the substrate. A precise O$_2$ pressure of $3.0-3.1\times10^{-5}$~mbar (with base pressure of the evaporator at $10^{-6}$~mbar) was selected through trial-and-error to ensure that the samples were in the regime of interest, close to the SIT transition, yet retaining insulating behavior down to temperatures of $\simeq 1.7$~K. Finally, the resist mask was removed using acetone (lift-off). Up to 20 samples were made for each batch, on a single 1~cm~$\times$~1~cm substrate. Different geometries were tested, with channel width $(W)\,\times$ length $(L)$ ranging 5~--~10~$\upmu$m~$\times$~20~--~25~$\upmu$m, except for certain samples designed as long wires (1~$\upmu$m~$\times$~25~$\upmu$m, see Table~\ref{table:treatments}). No external heating or cooling of the substrate was applied throughout these fabrication steps (evaporation and lift-off).

\subsection*{$\pmb{R_\square-T}$ characterization}
Following sample deposition, the DC electrical resistance $R$ of the samples was measured with a 4-contact method using a Keithley 6221 current source and a Keithley 2182A nanovoltmeter. The sheet resistance of the samples $R_\square = R\cdot L/W$ was then obtained from the actual geometric factor $L/W$ of samples, calculated from optical microscopy images. In the $R_\square$ range of our films (around 3~k$\Omega$ at room temperature), a small resistance drift was observed, well described by a $\log(t)$ decrease, where $t$ is the time elapsed at room temperature after deposition. To mitigate the effects of this drift, the samples were left at room temperature for at least one day after lift-off, prior to low temperature characterization and subsequent e-beam treatment. After one day, the relative variations in resistance were less than 1\% over the following day.
The samples were then cooled and measured down to $1.6-1.7$~K in a Variable Temperature Insert (VTI), in order to verify their insulating nature. For each sample, sufficiently low currents were selected to prevent self-heating while maintaining a high signal-to-noise ratio. A custom-made voltage preamp was occasionally employed to reduce the noise level of the most resistive samples.

Our deposition setup facilitated the achievement of high sample-to-sample repeatability (within the same batch deposited on a single 1~cm~$\times$~1~cm substrate) of both the geometrical factors of the a-InO$_x$ channel and their sheet resistance at both room temperature and $4.2$~K, with the exception of the long wires, for which the geometrical factors were less accurately determined during development. However, the batch-to-batch variation of $R_\square$ was between 2.7 and 3.7~k$\Omega$ at 295~K; 12 and 22~k$\Omega$ at 4.2~K; and 40~k$\Omega$ and 200~k$\Omega$ at 1.7~K (with 10 different evaporation runs conducted), despite maintaining identical nominal evaporation parameters (oxygen pressure and evaporation rate). This highlights the challenge of finely controlling the electrical properties of the a-InO$_x$ films solely through e-beam evaporation.

\subsection*{Sample treatment}
Following the initial characterization, the 1~cm~$\times$~1~cm substrates were transferred into the chamber of a LEO1530 scanning electron microscope, where they underwent exposure to the e-beam using a standard lithography controller and software (RAITH Elphy Plus). This system offers flexibility in adjusting both the acceleration voltage of the electron beam and the electron dose per surface area received by the samples, while also enabling precise control over which part of the channel is exposed (as detailed below). A reference sample from each substrate was kept unexposed throughout the experiments. This reference sample served to monitor any potential resistance drift at room temperature and to investigate possible spurious effects that could induce annealing. Subsequently, the samples were reinstalled in the VTI, and resistance-versus-temperature curves for both the e-beam-exposed and reference channels were measured and compared. No significant effects were observed on the reference samples following the exposure of the samples of the same batch, except for the natural evolution of a-InO$_x$ film properties due to the brief periods of time during which the samples remained at room temperature. As shown in Fig.~\ref{Figure2}(e), no morphological changes could be observed between untreated and treated channels in SEM micrographs.

\section*{Results}
Our main contribution is the proposal of a novel technique for local modulation of the electronic properties of a-InO$_x$ films across the superconductor-insulator transition through prolonged exposure to an e-beam. As shown in Fig.~\ref{Figure2}, the $R_\square$ versus $T$ curves demonstrate an evolution with increased dose (expressed as a multiple of the standard dose for PMMA resist, 250~$\upmu$C~cm$^{-2}$), from insulating ($dR/dT<0$ at low $T$) to metallic ($dR/dT>0$) and even superconducting behavior ($R_\square=0$) at sufficiently low temperatures. Defining $T_c$ from the linear extrapolation to $R = 0$ of the $R(T)$ curve at the point of maximum slope of the superconducting transition, a saturation value of $T_c\simeq2.85$~K was achieved with error on the reported $T_c$ below 0.05~K. For samples with gaps, the $T_c$ values were estimated by setting the zero of resistivity at the minimum resistivity measured. The evolution towards the superconducting state and the increase of $T_c$ are accompanied by a decrease of the normal state resistivity with the applied dose. Table~\ref{table:treatments} compiles all the treatments performed on different samples.

\begin{figure}[t]
    \centering
    \includegraphics[width=0.64\textwidth]{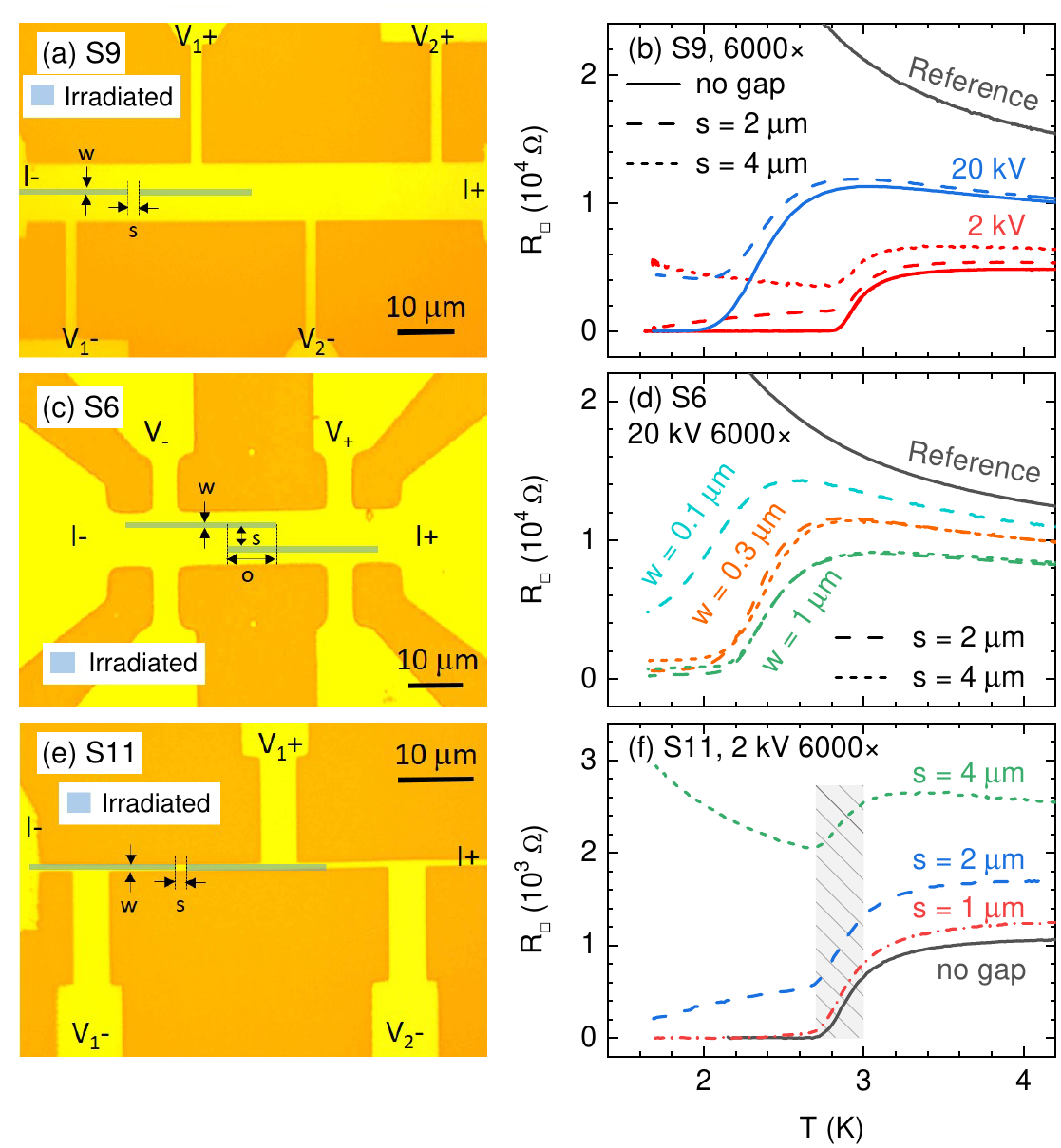}
    \caption{The spatial resolution of the e-beam treatment was tested using different acceleration voltages and geometries, and with a dose of 6000$\times$, almost sufficient to saturate the $T_c$ value (see Fig.~\ref{Figure2}~(d)). (a,~b)~$1\;\upmu$m wide lines, embedded in an insulating matrix, and separated by an insulating gap of length $s$, were drawn on substrate S9 at both 2~kV and 20~kV. (c,~d)~On substrate S6, lines of different width $w$, overlapping $10~\upmu$m longitudinally and separated by a distance $s$ transversally were drawn instead. (e,~f)~Substrate S11 was finally used to draw $1~\upmu$m wide fully superconducting channels separated by an insulating gap of length $s$. As shown, fully superconducting behavior (i.e., $R=0$) was not observed above 1.6~K for $s\geq2\;\upmu$m, and the gap remained insulating for $s\geq4\;\upmu$m. Nevertheless, the superconducting transition of the exposed part remained visible in all samples, evidencing the characteristic sudden resistance drop at a remarkably repeatable $T_c$ within the same batch [see e.g. the shaded area in~(f), with the exception of the 100~nm lines in~(d)].}\label{Figure4}
\end{figure}

\begin{table}[htb!]
\begin{tabular}{@{}cccccccc@{}}
\toprule
& \multirow{2}{*}{$\mathbf{V}$ \textbf{(kV)}} & \multirow{2}{*}{$\mathbf{I}$ \textbf{(nA)}} & \textbf{Dose} & \textbf{Channel Size} & \textbf{Exposed Size} & \multirow{2}{*}{$\mathbf{T_c}$ \textbf{(K)}} & \multirow{2}{*}{\textbf{Notes}} \\ 
& & & \textbf{($\times$250 $\mathbf{\upmu}$C/cm$^{2}$)} & \textbf{($\mathbf{\upmu}$m $\times$ $\mathbf{\upmu}$m)} & \textbf{(see main text)}  & & \textbf{} \\ \midrule
\multirow{3}{*}{\textbf{S2}}  & 20   & 1.04 & 10$\times$    & 5 $\times$ 25 & Whole channel & & \multirow{3}{*}{Fig. 2(a, c)}  \\
                              & 20   & 1.04 & 100$\times$   & 5 $\times$ 25 & Whole channel & $<1.6$ &  \\
                              & 20   & 4.2  & 1000$\times$  & 5 $\times$ 25 & Whole channel & 1.9 &\\ \midrule
\multirow{2}{*}{\textbf{S5}}  & 20   & 4.16 & 6000$\times$  & 10 $\times$ 25 & w=1 $\upmu$m & 2.65 &\multirow{2}{*}{$\Delta$T discussion} \\
                              & 20   & 1.03 & 6000$\times$  & 10 $\times$ 25 & w=1 $\upmu$m & 2.55 &\\ \midrule
\multirow{5}{*}{\textbf{S6}}  & 20   & 4.16 & 6000$\times$  & 10 $\times$ 25 & w=1 $\upmu$m, s=2 $\upmu$m, o=10 $\upmu$m   & $\lesssim2.21$ & \multirow{5}{*}{Fig. 4(c, d)} \\
                              & 20   & 4.16 & 6000$\times$  & 10 $\times$ 25 & w=0.3 $\upmu$m, s=2 $\upmu$m, o=10 $\upmu$m & $\lesssim2.16$ &  \\
                              & 20   & 1.03 & 6000$\times$  & 10 $\times$ 25 & w=0.1 $\upmu$m, s=2 $\upmu$m, o=10 $\upmu$m & $\lesssim1.80$ &  \\
                              & 20   & 4.16 & 6000$\times$  & 10 $\times$ 25 & w=1 $\upmu$m, s=4 $\upmu$m, o=10 $\upmu$m   & $\lesssim2.21$ &  \\
                              & 20   & 4.16 & 6000$\times$  & 10 $\times$ 25 & w=0.3 $\upmu$m, s=4 $\upmu$m, o=10 $\upmu$m & $\lesssim2.13$ &  \\ \midrule
\multirow{3}{*}{\textbf{S7}}  & 20   & 3.62 & 6000$\times$  & 10 $\times$ 25 & w=1 $\upmu$m & 1.96 &\multirow{3}{*}{Fig. 3c} \\
                              & 10   & 2.31 & 6000$\times$  & 10 $\times$ 25 & w=1 $\upmu$m & 2.36 &\\
                              & 5    & 1.74 & 6000$\times$  & 10 $\times$ 25 & w=1 $\upmu$m & 2.45 &\\ \midrule
\multirow{5}{*}{\textbf{S8}}  & 20   & 3.43 & 6000$\times$  & 10 $\times$ 20 & w=1 $\upmu$m & 1.93 &\multirow{5}{*}{Fig. 3c} \\
                              & 5    & 1.65 & 6000$\times$  & 10 $\times$ 20 & w=1 $\upmu$m & 2.58 &\\
                              & 2    & 1.32 & 6000$\times$  & 10 $\times$ 20 & w=1 $\upmu$m & 2.66 &\\
                              & 5    & 1.65 & 1000$\times$  & 10 $\times$ 20 & w=1 $\upmu$m & 1.53 &\\
                              & 2    & 1.32 & 1000$\times$  & 10 $\times$ 20 & w=1 $\upmu$m & 1.76 &\\ \midrule
\multirow{5}{*}{\textbf{S9}}  & 20   & 3.43 & 6000$\times$  & 10 $\times$ 20 & w=1 $\upmu$m & 2.10 &\multirow{5}{*}{\parbox{2 cm}{Fig.~\ref{Figure3}~(a~--~c)\hfill, Fig.~\ref{Figure4}~(a,~b)}} \\
                              & 2    & 1.31 & 6000$\times$  & 10 $\times$ 20 & w=1 $\upmu$m & 2.84 &\\
                              & 20   & 3.75 & 6000$\times$  & 10 $\times$ 20 & w=1 $\upmu$m, s=2 $\upmu$m & $\lesssim2.14$ &  \\
                              & 2    & 1.46 & 6000$\times$  & 10 $\times$ 20 & w=1 $\upmu$m, s=2 $\upmu$m & $\lesssim2.85$ &  \\
                              & 2    & 1.46 & 6000$\times$  & 10 $\times$ 20 & w=1 $\upmu$m, s=4 $\upmu$m & $\lesssim2.85$ &  \\ \midrule
\multirow{8}{*}{\textbf{S11}} & 2    & 1.48 & 2000$\times$  & 1 $\times$ 20 & Whole channel & 2.28 &\multirow{8}{*}{\parbox{2 cm}{Fig.~\ref{Figure2}~(b,~d)\hfill, Fig.~\ref{Figure4} (e,~f)}\hfill} \\
                              & 2    & 1.48 & 4000$\times$  & 1 $\times$ 20 & Whole channel & 2.60 & \\
                              & 2    & 1.48 & 6000$\times$  & 1 $\times$ 20 & Whole channel & 2.74 & \\
                              & 2    & 1.48 & 9000$\times$  & 1 $\times$ 20 & Whole channel & 2.83 & \\
                              & 2    & 1.48 & 12000$\times$ & 1 $\times$ 20 & Whole channel & 2.84 & \\
                              & 2    & 1.48 & 6000$\times$  & 1 $\times$ 20 & s=1 $\upmu$m & 2.71 &  \\
                              & 2    & 1.48 & 6000$\times$  & 1 $\times$ 20 & s=2 $\upmu$m & $\lesssim2.71$ &  \\
                              & 2    & 1.48 & 6000$\times$  & 1 $\times$ 20 & s=4 $\upmu$m & $\lesssim2.72$ &  \\ \midrule
\end{tabular}
\caption{E-beam treatments performed on the samples of this study. For samples with gaps, the $T_c$ values were estimated by setting the zero of resistivity at the minimum resistivity measured.}
\label{table:treatments}
\end{table}

Given this result, we tested the possibility of drawing superconducting microstructures within an insulating a-InO$_x$ matrix. We exposed lines measuring 1~$\upmu$m wide within $5-10\;\upmu$m wide channels (see Fig.~\ref{Figure3} (a)). A zero-resistance state was observed at sufficiently low temperatures, evidencing the presence of a continuous line of superconducting material in the channel. The effect of the acceleration voltage on $T_c$ is clearly discernible: for S9 (Fig.~\ref{Figure3} (b), and squares and solid green line in (c)), we observed a $T_c$ increase from 2.1~K up to 2.84~K as the acceleration voltage was decreased from 20 to 2~kV. This trend was consistently observed across different samples, as shown in Fig.~\ref{Figure3}~(c). A rough estimate of the effective e-beam-treated line width can be obtained from a comparison of the $R_\square$ values in the normal state ($T\simeq 4.2$~K) before and after the e-beam treatment. For a $1~\upmu$m wide line exposed in a 10~$\upmu$m wide channel (corresponding to S9), $R_\square$ decreased from 15~k$\Omega$ to 5~k$\Omega$. In other samples where the whole channel was exposed, $R_\square$ was approximately $\simeq1$~k$\Omega$ for similar e-beam parameters (acceleration voltage of 2~kV and dose factor of 6000$\times$). Assuming that the measured resistance results from a parallel conduction between a homogeneous line affected by the e-beam and the rest of the channel, which remains unaffected, we infer that the effective line width is approximately $\sim1.4~\upmu$m, significantly larger than the nominal width.\\

Other microstructures were drawn to estimate the spatial resolution of the e-beam treatment. For instance, a single line of width $w$ much narrower than the total channel width was exposed while leaving a length $s$ unexposed in the center of the wire, creating an insulating interruption or ``gap'', as shown in Fig.~\ref{Figure4}~(a). Another geometry consisted in exposing two parallel wires offset by a distance $s$, but longitudinally overlapping by a distance $o$, as shown in Fig.~\ref{Figure4}~(c). Additionally, in samples shaped as long, thin wires, the entire channel width could be exposed at high doses while leaving a small insulating gap $s$ embedded in a fully superconducting wire, as pictured in Fig.~\ref{Figure4}~(e). The results of these tests were unequivocal -- we successfully created short insulating interruptions within the superconducting exposed sections, as evidenced by the $R_\square(T)$ curves shown in Fig.~\ref{Figure4}~(b,~d,~f). At lower temperatures than the clearly visible superconducting transition of the exposed parts, an insulating behavior ($dR/dT<0)$ was observed down to $s\gtrsim4\;\upmu$m. For gaps below this value, the wires become either metallic ($dR/dT>0$) or superconducting. This indicates that the resolution of our treatment is about $\sim2\;\upmu$m, half the size of the gap. A spatial resolution of this order of magnitude was observed for all acceleration voltages. Finally, for one of the batches (S6), we investigated the effect of reducing line widths by adopting the second geometry and employing $w = $~(100~nm, 300~nm, $1\;\upmu$m). In this case, we observed very similar $T_c$ and resolution length scale for $w=300$~nm and $1\;\upmu$m, while a smaller $T_c$ was obtained for $w=100$~nm (see Fig.~\ref{Figure4}~d). 

\section*{Discussion}

We will now discuss the mechanisms that could explain the observed drop in electrical resistance after e-beam exposure. A standard process for tuning amorphous InO$_x$ film resistance across the SIT is annealing \cite{ShaharPRB1992, SambandamurthyPRL2004}. Below $\simeq 100^\circ$C , annealing induces an irreversible decrease in resistance while preserving the amorphous structure of the films \cite{OvadyahuJPC1986}, and is accompanied by a densification of the film \cite{OvadyahuPRB2017Slow, OvadyahuPSS2020} while the charge carrier density remains essentially unchanged \cite{ShaharPRB1992, SambandamurthyPRL2004}. In films with low carrier density ($n\simeq 10^{20}$~cm$^{-3}$), which exhibit high resistance in their non-annealed state \cite{OvadyahuPRB2017Memory}, the resistance drop at room temperature can be of several orders of magnitude \cite{OvadyahuPRB2017Memory}. However, in our amorphous films, which are less resistive and have higher charge carrier density (according to our e-beam evaporation parameters -- evaporation rate of 2~\AA/s and $P_{\rm O_2}=3\times 10^{-5}$~mbar, $n \sim10^{22}$~cm$^{-3}$), annealing may decrease the room temperature resistance by at most a factor of 4 \cite{OvadyahuPSS2020}. Remarkably, this decrease is commensurate  with the decrease observed at room temperature between the resistance of an unexposed film, and the same film exposed to the highest doses (12000$\times$). For example, in S11, $R_\square$ decreases from 2.6~k$\Omega$ prior to e-beam exposure to 0.6~k$\Omega$ after (note that the relative changes at 4.2~K are larger: from 15~k$\Omega$ to 0.7~k$\Omega$). Hence, the question arises as to whether the SEM e-beam can induce such annealing. When the electron probe diameter is much smaller than the electron range $\lambda$ in a solid, which seems reasonable in our case (see simulations below), an estimate of the local temperature increase $\Delta T$ is given by  \cite{ReimerBook1998}
\begin{equation}\label{Equation2}
    \Delta T = 3fUI/(2\pi\kappa\lambda)\;.
\end{equation}
In Eq.~\ref{Equation2}, $f$ is the power conversion factor to heat, $U$ the acceleration voltage of the electrons, $I$ the current intensity of the e-beam, $\lambda$ the electron range and $\kappa$ the thermal conductivity of the material. Taking $f = 100\%$, $U=20$~kV, $I=4$~nA, $\lambda \simeq 1\;\upmu$m and $\kappa \simeq 100$~W$\,$m$^{-1}$K$^{-1}$ (for Si++ substrate), we obtain $\Delta T \simeq 0.4^\circ$C. Taking the values for $U=2$~kV, $I=1.5$~nA and $\lambda=10$~nm instead, we obtain $\Delta T\simeq1.5$~K. These values are significantly smaller than the $\Delta T$ required for annealing of amorphous InO$_x$, which is typically several tens of degrees. It is worth noting that much larger $\Delta T$ over 1000~K have been reported \cite{WangNRL2019} when using higher voltages in highly focused TEM (200~kV) or higher currents in a 40 kV SEM ($\sim$10~mA), while for similar voltages and doses using a SEM in similar conditions, the $\Delta T$ observed is in any case below 5~K. To evaluate the significance of an e-beam heating effect, we conducted experiments comparing resistance changes obtained under 20~kV but using two different currents and the same dose: 4.2~nA (120~$\upmu$m diaphragm) and 1.0~nA (60~$\upmu$m diaphragm). Minute $R_\square$ and $T_c$ changes were observed: $R_\square$ at room temperature increased from 2.25~k$\Omega$ to 2.28~k$\Omega$, and $T_c$ decreased from 2.65~K to 2.55~K.

\begin{figure}[t]
    \centering
    \includegraphics[width=\textwidth]{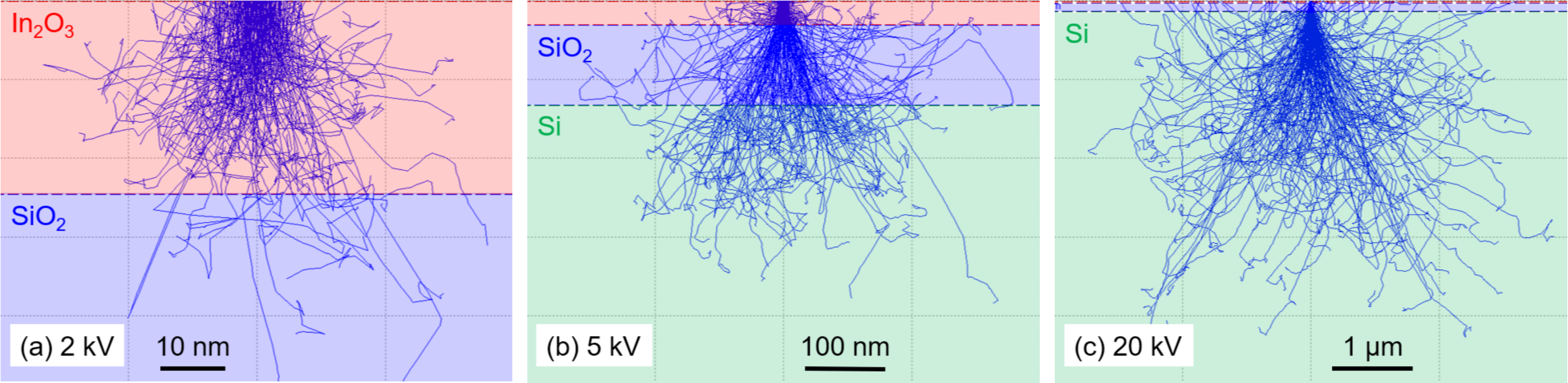}
    \caption{Simulations of the e-beam electron trajectories at different acceleration voltages: (a) 2~kV; (b) 5~kV; (c) 20~kV. The simulations were done using CASINO \cite{CASINO}, and assuming 30~nm of In$_2$O$_3$ (red), 100~nm of SiO$_2$ (blue) and 10~$\upmu$m of Si (green).}
    \label{Figure5}
\end{figure}

The simulations of the electron trajectories at different acceleration voltages can provide insight into the potential mechanisms underlying the observed resistance changes. To conduct these simulations, we used CASINO Monte-Carlo software \cite{CASINO} and considered a structure comprising 30~nm of In$_2$O$_3$, 100 nm of SiO$_2$, and 10 $\upmu$m of Si. From the results, illustrated in Fig.~\ref{Figure5}, we conclude that, at 20~kV, the electron range extends to a few micrometers within the Si substrate, whereas at 2~kV it remains confined within the 30~nm layer of a-InO$_x$. However, the part of the a-InO$_x$ film directly affected by primary (and thus secondary) electrons is confined to a few tens of nanometers radially from the electron beam (assumed to have a diameter of 10~nm). Given that the region with reduced resistance due to the e-beam treatment extends approximately 1~$\upmu$m beyond the exposed area, direct action of the e-beam electrons such as an e-beam-induced chemical reduction, previously reported in vanadium and titatium oxides \cite{Kern2006,Kern2006b, Zhang2023, Zhang2024}, cannot explain the observed changes. Nevertheless, the effect of UV exposure on indium oxide electrical resistance is well established. A conducting state was found in previous work on UV exposure in amorphous InO$_x$ films, observing significant resistance changes which were attributed to a UV-photoreduction mechanism, where the conducting photoreduced state was found to be unstable in oxidizing atmosphere; subsequent ozone or O$_2$ plasma exposure of the films achieved the opposite effect, reverting the samples to the high resistance state \cite{PashmakovJNCS1993, PashmakovSSC1993, FritzscheSEM1994, ClaflinJEM1996}. These photoreduction-oxidation cycles were associated with significant changes in the oxygen content of the topmost 10~nm layer of the film, and correlated with oxygen outdiffusion from the film \cite{ClaflinJEM1996}. Recent work by Lo Mastro et al.~\cite{LoMastro2024} studied the ageing of In$_2$O$_3$ films after different treatments to increase its carrier density, such as ion implantation and UV photoreduction, concluding that \emph{“UV photo-reduction, both in argon saturated atmosphere or in air, is a simpler and very effective tool to induce intrinsic doping by formation of oxygen vacancies.”}. In our case, a fraction of the radiation emitted during electron–matter interactions in pear-shaped volume of interaction (see simulations) extends into the UV range, as the continuous X-ray spectrum reaches down to visible and near-UV energies \cite{ReimerBook1998}. In addition, cathodoluminescence may contribute as another UV source: nanocrystalline indium oxide and indium tin oxide films show a cathodoluminescence peak in the visible, extending into the near-UV \cite{Korotcenkov2007, Maestre2008}. Although our films are amorphous rather than nanocrystalline, the underlying mechanism attributed to these emissions -- transitions between shallow energy levels associated with point defects -- could also play a role. While the exact microscopic origin of the resistance decrease remains to be established, these considerations indicate that indirect radiative processes are a plausible contributor to the observed behaviour.

Additional tests were conducted to evaluate the stability of the low resistance state obtained after e-beam exposure. A superconducting channel (with $T_c$ around 2~K) was exposed to ambient air at room temperature for 3.5 days. A small resistance increase of the exposed channel from 2.66 to 2.70~k$\Omega$ at 300~K was observed, while during the same time, the resistance of a non-exposed insulating channel decreases by a few \%. Additionally, a small increase of $T_c$ was noted, from 1.91~K to 1.96~K. These observations indicate that the exposed channel is not reverting to the non-exposed insulating state. Note that for samples kept at lower temperature ($-30^\circ$C), no significant evolution was observed. This room temperature stability rules out a surface-limited mechanism driven by oxygen exchange between the film and the atmosphere. Mechanisms not involving oxygen outdiffusion which can affect the bulk of the films have been previously discussed in a-InO$_x$ films \cite{PashmakovJNCS1993}. If a photon causes an electron transfer from O$^{2-}$ to In$^{3+}$ ion, the resulting O$^{-}$ ion may form weak covalent bonds with a neighboring O$^{2-}$ ion or stronger bonds with another O$^-$ ion. This process increases the electron concentration, leading to higher charge carrier density and electrical conductance.

\begin{figure}[t]
    \centering
    \includegraphics[width=0.8\textwidth]{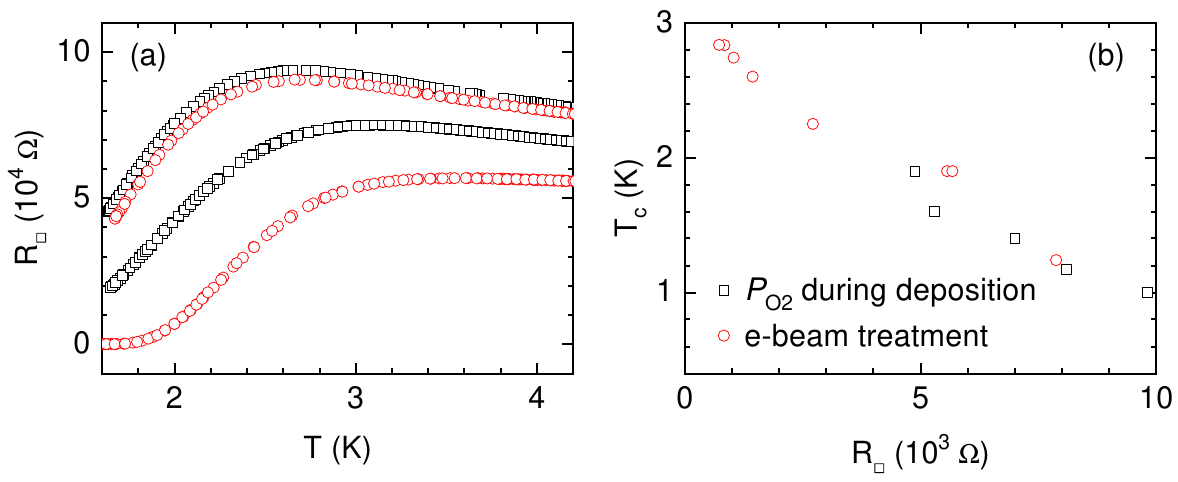}
    \caption{Comparison between samples manufactured using the standard method (changing $P_{O_2}$ during deposition, black squares) and the technique proposed in this paper (red circles). (a) $R_\square$ versus $T$ curves between 1.7~K and 4.2~K for samples obtained using the standard method (upper curve: 40~nm thick, $P_{O_2}=3\times10^{-5}$~mbar; lower curve: 30~nm,~no~O$_2$) and samples from this paper. (b) $T_c$ versus $R_\square$ at 4.2~K for samples obtained using the standard method (in ascending order of $T_c$: 20~nm,~no~O$_2$; 40~nm, $P_{O_2}=3\times 10^{-5}$~mbar; 30~nm,~no~O$_2$; 70~nm, $P_{O_2}=3.8\times 10^{-5}$~mbar; 47~nm, $P_{O_2}=3.5\times 10^{-5}$~mbar) and samples from this paper.}
    \label{Figure6}
\end{figure}

Another important question pertains to how the $R_\square-T$ curves of our e-beam-exposed superconducting films compare to those obtained through the conventional method of adjusting the oxygen pressure during film deposition. We prepared a few films with thicknesses ranging from 20 to 47~nm and oxygen pressures between 0 (no added oxygen in the evaporation chamber) and $6\times 10^{-5}$~mbar with a fixed evaporation rate of 2~\AA/s. In these films, the $R_\square$ value for a given thickness is determined by the amount of oxygen vacancies present \cite{OvadyahuJPC1986, GivanPRB2012}. The highest $T_c$ achieved in 30~nm thick films was about 1.4~K, when no O$_2$ was introduced during e-beam evaporation. The $R_\square-T$ curve of this film appears to fall between those of exposed films with $T_c$ of 1.2 and 1.9~K (see Fig.~\ref{Figure6}). Moreover, comparing an e-beam-exposed sample and a 40~nm thick film evaporated under $3\times 10^{-5}$~mbar of O$_2$, both having similar $T_c\approx1.2$~K, their $R_\square-T$ curves are almost identical. These results suggest that the effect of e-beam exposure is equivalent to a change in oxygen vacancies content, at least regarding the $R_\square-T$ dependency. Additionally, our e-beam exposure process enables achieving much higher $T_c$ in very thin films compared with the standard evaporation process: a saturation value of $\sim2.85$~K for our 30~nm thick films, compared to $\sim$1.4~K in 30~nm thick unexposed films when no oxygen was added during the evaporation in our setup. However, it is worth mentioning that critical temperatures as high as 2.4~K have been achieved using different setups \cite{Crane2007,Liu2013,Wang2018}, while $T_c$ up to 3.4~K can be obtained in a-InO$_x$ films by annealing and increasing the thickness \cite{SacepePRB2015, Charpentier2024}. \\

\section*{Conclusion}

A new technique to locally transform an insulating amorphous indium oxide film into superconducting by e-beam exposure has been proposed. The $T_c$ values of the irradiated superconducting region can be finely tuned by adjusting the acceleration voltage and electron dose, achieving a maximum $T_c$ value of approximately 2.8~K in 30~nm thick films. The resulting $R_\square-T$ curves closely resemble those of amorphous indium oxide films deposited under different oxygen pressures.

This innovative approach to fabricating superconducting amorphous indium oxide films opens avenues for fundamental studies of the superconducting-insulating transition in this system. Typically, films made under different oxygen pressures lack precise control over $R_\square$ and $T_c$, necessitating subsequent annealing steps. Our SEM e-beam exposure method enables fine control of electrical parameters, allowing the production of samples with varied $R_\square$ and $T_c$ values within a single SEM session and after a single evaporation step, while also ensuring same-batch sample-to-sample reproducibility.

Moreover, the resolution achieved is approximately $\sim2\;\upmu$m, enabling the creation of superconducting lines just a few micrometers wide, with well-defined $T_c$ or with micrometer-scale changes in $T_c$, within an insulating matrix of amorphous indium oxide. These characteristics hold promise for various applications, such as the development of high kinetic inductances or superconducting nanowire single-photon detectors.

The observed resistance changes may be related to the photoreduction of the films by X-rays and/or UV radiation emitted during e-beam interactions with the film and its substrate. Further detailed structural and chemical investigations are required to elucidate this aspect. The relaxation of these materials at room temperature remains an open question, and its study could provide valuable insight into the mechanisms underlying this behaviour. \\

\section*{Data availability}
Data supporting the findings of this study are included within the article. The raw data underlying the results can be made available upon reasonable request. Requests for data should be directed to I. F. Llovo or J. Delahaye.

\bibliography{references.bib}

\section*{Acknowledgements}
The authors would like to thank Dr. T. Grenet and Prof. M. V. Ramallo for their invaluable insight during our conversations on the topic. I. F. Llovo acknowledges financial support from Xunta de Galicia through grant ED481A-2020/149. J. Delahaye acknowledges financial support from the Agence Nationale de la Recherche (grant ANR-19-CE30-0014-04).

\section*{Author contributions}
Both authors contributed to the conceptualization of the study, experimental work, analysis and writing of the manuscript.

\section*{Competing interests}
The authors declare no competing interests.

\end{document}